\documentstyle[12pt]{article}
\def\b{\beta}
\def\t{\theta}

\def\ch{ch}
\def\sh{sh}

\def\cosh{cosh}
\def\ve{\varepsilon}
\def\non{\nonumber}

\begin{document}

\newpage
\pagestyle{empty}
\begin{flushright}
DTP/95-23 \\
hep-th/9505138 \\
May 1995
\end{flushright}
\begin{center}
{\Large {\bf Boundary Reflection Matrix \\
for $D_4^{(1)}$ Affine Toda Field Theory}}
\end{center}
\begin{center}
{\bf J. D. Kim\footnote{jideog.kim@durham.ac.uk} \footnote{On leave
of absence from Korea Advanced Institute of Science and Technology} } \\
{\it Department of Mathematical Sciences, \\
University of Durham, Durham DH1 3LE, U.K.} \\
\vspace{0.5cm}
{\bf H. S. Cho} \\
{\it Institute for Theoretical Physics, \\
20, Marshall Terrace, Durham DH1 2HX, U.K.}  \\
\end{center}
\vspace{1.5cm}

\begin{center}
ABSTRACT
\end{center}

We present one loop boundary reflection matrix
for $d_4^{(1)}$ Toda field theory defined on a half line
with the Neumann boundary condition.
This result demonstrates a nontrivial cancellation of non-meromorphic
terms which are present when the model has a particle
spectrum with more than one mass.
Using this result, we determine uniquely the exact boundary
reflection matrix which turns out to be \lq non-minimal'
if we assume the strong-weak coupling \lq duality'.

\newpage
\pagestyle{plain}

\section*{I. Introduction}
The exact $S$-matrix for integrable quantum field theory defined
on a full line has been conjectured using the symmetry principles
such as Yang-Baxter equation, unitarity, crossing relation,
real analyticity and bootstrap equation\cite{ZZ,AFZ,BCDS,CM,DGZ}.
This program entirely relies on the assumed quantum integrability of the model
as well as the fundamental assumptions
such as strong-weak coupling \lq duality' and \lq minimality'.

In order to determine the exact $S$-matrix uniquely,
Feynman's perturbation theory has been used\cite{BCDS2,BS,CKK,BCKKS,SZ}
and shown to agree well with the conjectured \lq minimal' $S$-matrices.
In perturbation theory, $S$-matrix is extracted
from the four-point correlation function with LSZ reduction formalism.
Especially, the singularity structures were examined in terms of
Landau singularity\cite{ELOP}, of which odd order poles are interpreted
as coming from the intermediate bound states.

About a decade ago, integrable quantum field theory defined on a half line
$(-\infty < x \leq 0)$
was studied using symmetry principles under the assumption that
the integrability of the model remains intact\cite{Che}.
The boundary Yang-Baxter equation, unitarity relation
for boundary reflection matrix $K_a^b(\t)$
which is conceived to describe the scattering process off a wall
was introduced\cite{Che}.
Recently, boundary crossing relation\cite{GZ} and boundary bootstrap
equation\cite{FK} was introduced.
Subsequently, some exact boundary reflection matrices
have been conjectured\cite{GZ,FK,Gh,Sasaki,CDRS} for affine Toda field
theory(ATFT).

In order to determine the boundary reflection matrix uniquely,
we have developed a method\cite{Kim} in the framework
of the Lagrangian quantum field theory with a boundary\cite{Sym,DD,BM}.
The idea is to extract the boundary reflection matrix
directly from the two-point correlation function in the coordinate space.

Using this formalism, we determined the exact boundary reflection matrix
for sinh-Gordon model($a_1^{(1)}$ affine Toda theory)
and Bullough-Dodd model($a_2^{(2)}$ affine Toda theory)
with the Neunmann boundary condition
modulo \lq a universal mysterious factor half'.
If we assume the strong-weak coupling \lq duality', these solutions
are unique.
Above two models have a particle spectrum with only one mass.
On the other hand, when the theory has a particle spectrum with more
than one mass, each one loop contribution from different types
of Feynman diagrams has non-meromorphic terms.

In this paper, we evaluate one loop boundary reflection matrix for
$d_4^{(1)}$ affine Toda field theory and show a remarkable cancellation
of non-meromorphic terms among themselves.
This result also enables us to determine the exact
boundary reflection matrix uniquely under the assumption of
the strong-weak coupling \lq duality'.
The boundary reflection matrix has singularities which
can be accounted for by a new type of singularities
of Feynman diagrams for a theory defined on a half line.

In section II, we review the formalism developed in ref.\cite{Kim}.
In section III, we present one loop result for
$d_4^{(1)}$ affine Toda theory
and determine the exact boundary reflection matrix.
We also present the complete set of solutions of the boundary
bootstrap equations.
Finally, we make conclusions in section IV.

\section*{II. Boundary Reflection Matrix}
The action for affine Toda field theory defined on a half line
($-\infty < x \leq 0$) is given by
\begin{equation}
S(\Phi) = \int_{-\infty}^{0} dx \int_{-\infty}^{\infty} dt
\left ( \frac{1}{2}\partial_{\mu}\phi^{a}\partial^{\mu}\phi^{a}
-\frac{m^{2}}{\b^{2}}\sum_{i=0}^{r}n_{i}e^{\b \alpha_{i} \cdot \Phi}
\right ),
\end{equation}
where
\begin{displaymath}
\alpha_{0} = -\sum_{i=1}^{r}n_{i}\alpha_{i},~~ \mbox{and}~~ n_{0} = 1 .
\end{displaymath}
The field $\phi^{a}$ ($a=1,\cdots,r$) is $a$-th component of the scalar field
$\Phi$,
and $\alpha_{i}$ ($i=1,\cdots,r$) are simple roots of a Lie algebra
$g$ with rank $r$ normalized so that the universal function $B(\b)$
through which the dimensionless coupling
constant $\b$ appears in the $S$-matrix takes the following form:
\begin{equation}
B(\b)=\frac{1}{2\pi}\frac{\b^2}{(1+\b^2/4\pi)}.
\label{Bfunction}
\end{equation}
The $m$ sets the mass scale and the $n_i$s are the so-called Kac labels
which are characteristic integers defined for each Lie algebra.

Here we consider the model with no boundary potential,
which corresponds to the Neumann boundary condition:
$\frac{ \partial \phi^a} {\partial x} =0$ at $x=0$.
This case is believed to be quantum stable in the sense that
the existence of a boundary does not change
the structure of the spectrum.

In classical field theory, it is quite clear how we extract
the boundary reflection matrix. It is the coefficient of
reflection term in the two-point correlation function namely it is 1.
\begin{eqnarray}
G_N (t',x';t,x) &=& G(t',x';t,x) + G(t',x';t,-x) \\
             &=& \int \frac{d^2 p}{(2 \pi)^2} \frac{i}{p^2-m_a^2+i \ve}
  e^{-i w (t'-t)}
 (  e^{i k (x'-x)} +  e^{i k (x'+x)} ).  \non
\end{eqnarray}
We may use the $k$-integrated version.
\begin{equation}
G_N (t',x';t,x) = \int \frac{d w}{2 \pi} \frac{1}{2 \bar{k}}
  e^{-i w (t'-t)} (  e^{ i \bar{k} |x'-x| } +  e^{-i \bar{k} (x'+x)} ),
{}~~ \bar{k}=\sqrt{w^2-m_a^2}.
\end{equation}
We find that the unintegrated version is very useful to extract the
asymptotic part of the two-point correlation function far away from the
boundary.

In quantum field theory, it also seems quite natural to extend above
idea in order to extract the quantum boundary reflection matrix directly
from the quantum two-point correlation function.
This idea has been pursued in ref.\cite{Kim} to extract one loop
boundary reflection matrix.

To compute two-point correlation functions at one loop order,
we follow the idea of the conventional perturbation theory\cite{Sym,DD,BM}.
That is, we generate relevant Feynman diagrams
and then evaluate each of them by using the zero-th order
two-point function for each line occurring in the Feynman diagrams.

At one loop order, there are three types of Feynman diagram contributing to
the two-point correlation function as depicted in Figure 1.

\linethickness{0.5pt}
\begin{picture}(350,175)(-50,-110)
\put(0,0){\circle{40}}
\put(-20,50){\line(0,-1){100}}
\put(-15,35){a}
\put(-15,-35){a}
\put(25,0){b}
\put(-18,-70){Type I.}
\put(170,0){\circle{40}}
\put(130,0){\line(1,0){20}}
\put(130,50){\line(0,-1){100}}
\put(135,35){a}
\put(135,-35){a}
\put(137,5){b}
\put(195,0){c}
\put(135,-70){Type II.}
\put(300,0){\circle{40}}
\put(300,20){\line(0,1){30}}
\put(300,-20){\line(0,-1){30}}
\put(305,35){a}
\put(305,-35){a}
\put(270,0){b}
\put(325,0){c}
\put(280,-70){Type III.}
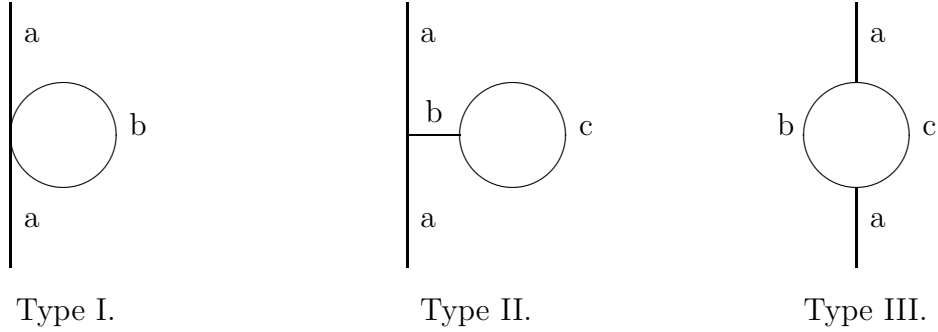
\put(10,-100){Figure 1. Diagrams for the one loop two-point function.}
\end{picture}

For a theory defined on a full line which has translational symmetry
in space and time direction,
Type I, II diagrams have logarithmic infinity
independent of the external energy-momenta
and are the only divergent diagrams in 1+1 dimensions.
This infinity is usually absorbed into
the infinite mass renormalization.
Type III diagrams have finite corrections
depending on the external energy-momenta
and produces a double pole to the two-point correlation function.

The remedy for these double poles is to introduce a counter term
to the original Lagrangian to cancel this term(or to renormalize the mass).
In addition, to maintain the residue of the pole, we have to
introduce wave function renormalization.
Then the renormalized two-point correlation function remains the same
as the tree level one with renormalized mass $m_a$,
whose ratios are the same as the classical value.
This mass renormalization procedure can be generalized
to arbitrary order of loops.

Now let us consider each diagram for a theory defined on a half line.
Type I diagram gives the following contribution:
\begin{equation}
\int_{-\infty}^{0} d x_1  \int_{-\infty}^{\infty} d t_1
 G_N (t,x;t_1,x_1) ~ G_N (t',x';t_1,x_1) ~ G_N (t_1,x_1;t_1,x_1).
\label{TypeI}
\end{equation}
{}From Type II diagram, we can read off the following expression:
\begin{equation}
\int_{-\infty}^{0} d x_1 d x_2 \int_{-\infty}^{\infty} d t_1 d t_2
 G_N (t,x;t_1,x_1) ~ G_N (t',x';t_1,x_1) ~ G_N (t_1,x_1;t_2,x_2)
\end{equation}
\begin{displaymath}
{}~~~~~~~~~~ G_N (t_2,x_2;t_2,x_2).
\end{displaymath}
Type III diagram gives the following contribution:
\begin{equation}
\int_{-\infty}^{0} d x_1  d x_2 \int_{-\infty}^{\infty} d t_1 d t_2
 G_N (t,x;t_1,x_1) ~ G_N (t',x';t_2,x_2) ~ G_N (t_2,x_2;t_1,x_1)
\end{equation}
\begin{displaymath}
G_N (t_2,x_2;t_1,x_1).
\end{displaymath}

After the infinite as well as finite mass renormalization,
the remaining terms coming from type I,II and III diagrams
can be written as follows with different $I_i$ functions:
\begin{equation}
\int \frac{dw}{2 \pi} \frac{dk}{2 \pi} \frac{dk'}{2 \pi}
   e^{-iw(t'-t)}  e^{i (kx+k'x')}
 \frac{i}{w^2-k^2-m_a^2+i \ve} \frac{i}{w^2-k'^2-m_a^2+i \ve}
 I_i(w,k,k').
\label{General}
\end{equation}
Contrary to the other terms which resemble those of a full line,
this integral has two spatial momentum integration.

In the asymptotic region far away from the boundary,
these terms can be evaluated up to exponentially damped
term as $x, x'$ go to $-\infty$, yielding the following result
for the elastic boundary reflection matrix $K_a(\t)$ defined
as the coefficient of the reflected term of the two-point
correlation function.
\begin{equation}
\int \frac{dw}{2 \pi} e^{-iw(t'-t)} \frac{1}{2 \bar{k}}
 (  e^{i \bar{k} |x'-x|} +K_a(w)  e^{-i \bar{k} (x'+x)} ),
  ~~ \bar{k}=\sqrt{w^2-m_a^2}.
\end{equation}
$K_a(\t)$ is obtained using $w=m_a \cosh\t$.

Here we list each one loop contribution to $K_a(\t)$
from the three types of diagram depicted in Figure 1\cite{Kim}:
\begin{eqnarray}
K_a^{(I)}(\t) &=& \frac{1}{4 m_a \sh\t} ( \frac{1}{2 \sqrt{m_a^2
\sh^2\t+m_b^2}}
    +\frac{1}{2 m_b} ) ~C_1 ~S_1,  \\
K_a^{(II)}(\t) &=& \frac{1}{4 m_a \sh\t}
  ( \frac{-i}{ (4 m_a^2 \sh^2\t +m_b^2) 2 \sqrt{m_a^2 \sh^2\t+m_c^2}}
    +\frac{-i}{ 2 m_b^2 m_c} ) ~C_2 ~S_2,  \\
K_a^{(III)}(\t) &=& \frac{1}{4 m_a \sh\t}
  ( 4 I_3(k_1=0,k_2=\bar{k} )+4 I_3(k_1=\bar{k} ,k_2=0) ) ~ C_3 ~S_3,
\label{K-III}
\end{eqnarray}
where \lq a universal mysterious factor half' is included.
$C_i, S_i$ denote numerical coupling factors and symmetry factors,
respectively. $I_3$ is defined by
\begin{equation}
I_3 \equiv
  \frac{1}{4}
 (\frac{i}{2 \bar{w}_1 (\bar{w}_1-\tilde{w}_1^+) (\bar{w}_1-\tilde{w}_1^-)}
  + \frac{i}{(\tilde{w}_1^+ -\bar{w}_1) (\tilde{w}_1^+ +\bar{w}_1)
        (\tilde{w}_1^+ -\tilde{w}_1^- )  } ),
\end{equation}
where
\begin{eqnarray}
\bar{w}_1=\sqrt{k_1^2+m_b^2}, &
\tilde{w}_1^+ =w+\sqrt{k_2^2+m_c^2}, & \tilde{w}_1^- =w-\sqrt{k_2^2+m_c^2}.
\end{eqnarray}
It should be remarked that this term should be symmetrized with respect
to $m_b, m_c$ with a half.

Let us remark a few interesting points.
Firstly, above expressions have non-meromorphic terms when
the theory has a mass spectrum with more than one mass.
Secondly, they have singularities which are absent for the
same Feynman diagrams from the theory on a full line.
Later, we will see a nontrivial cancellation of non-meromorphic
terms and the fact that the new type of singularities accounts for the
singularities of the exact boundary reflection matrix.

\section*{III. $d_4^{(1)}$ affine Toda theory}
We have to fix the normalization of roots so
that the standard $B(\b)$ function takes the form given in
Eq.(\ref{Bfunction}).

We use the Lagrangian density given as follows.
\begin{equation}
{\cal L} = \sum_{i=1}^{4} \frac{1}{2} \partial_\mu \phi_{i}
 \partial^\mu \phi_{i} -V(\phi),
\end{equation}
\begin{eqnarray}
V(\phi) &=& \frac{1}{2} m^2
  (2 \phi_{1}^{2} +6\phi_{2}^{2}+ 2\phi_{3}^{2}+ 2\phi_{4}^{2}) \non  \\
 &+&  \frac{1}{\sqrt{2}} m^2\b (-\phi_{1}^{2}\phi_{2}-\phi_{3}^{2}\phi_{2}
-\phi_{4}^{2}\phi_{2}+\phi_{2}^{3} -2\phi_{1}\phi_{3}\phi_{4})  \non \\
 &+&  \frac{1}{24} m^2 \b^2 (\phi_{1}^{4} + \phi_{3}^{4} +\phi_{4}^{4}
+9\phi_{2}^{4}+6\phi_{1}^{2}\phi_{2}^{2}+6\phi_{1}^{2}\phi_{3}^{2}
+6\phi_{1}^{2}\phi_{4}^{2} \non \\
 & & ~~~~~~~~~  +  6\phi_{2}^{2}\phi_{3}^{2} + 6\phi_{2}^{2}\phi_{4}^{2}
    + 6\phi_{3}^{2}\phi_{4}^{2} +24\phi_{1}\phi_{2}\phi_{3}\phi_{4})
    + O(\b^3).    \non
\label{VV}
\end{eqnarray}
The scattering matrix of this model is given by the following\cite{BCDS}.
\begin{eqnarray}
S_{11}(\t)=S_{33}=S_{44}=\{1\} \{5\}, & S_{22}=\{1\} \{5\} \{3\} \{3\}, \\
S_{12}(\t)=S_{32}=S_{42}=\{2\} \{4\}, & S_{13}=S_{14}=S_{34}=\{3\},  \non
\end{eqnarray}
\begin{displaymath}
  \{ x \}=\frac{(x-1) (x+1)}{(x-1+B) (x+1-B)},
  ~~~~ (x) = \frac{ \sh( \t /2 + i \pi x /2 h )}
            { \sh( \t /2 - i \pi x /2 h )}.
\label{Blockx}
\end{displaymath}
Here $B$ is the same function defined in Eq.(\ref{Bfunction}).
For this model, $h=6$ and from now on we set $m=1$.
Due to the triality symmetry among $\phi_1$, $\phi_3$ and $\phi_4$,
we have only to consider one of the light particles
and the heavy particle. We choose $\phi_1$ and $\phi_2$.

First, we consider the light particle.
For type I diagram, there are four possible configurations
three of which yield identical contribution.
We follow the notation of Figure 1.
For $b= \phi_1, \phi_3$ and $\phi_4$,
\begin{equation}
K_1^{(I-1)}= \frac{1}{4 \sqrt{2} \sh\t} ( \frac{1}{2 \sqrt{2} \ch\t}
 +\frac{1}{2 \sqrt{2} })
    \times (-i \b^2) \times \frac{1}{2} \times 3.
\end{equation}
For $b=\phi_2$,
\begin{equation}
K_1^{(I-2)}= \frac{1}{4 \sqrt{2} \sh\t} (\frac{1}{2 \sqrt{2 sh^2 \t+6} }
 +\frac{1}{2 \sqrt{6} })
    \times (\frac{-i}{4} \b^2) \times 2.
\end{equation}

For type II diagram, there are also four possible configurations
three of which yield identical contribution.
For $b=\phi_2$, $c=\phi_1, \phi_3$ and $\phi_4$,
\begin{equation}
K_1^{(II-1)}= \frac{1}{4 \sqrt{2} \sh\t}  (\frac{1}{(8 \sh^2\t +6)}
 \frac{-i}{2 \sqrt{2} \ch\t} +\frac{-i}{12 \sqrt{2} })
 \times (\frac{-1}{2} \b^2) \times 2 \times 3.
\end{equation}
For $b=c=\phi_2$,
\begin{equation}
K_1^{(II-2)}= \frac{1}{4 \sqrt{2} \sh\t}  (\frac{1}{(8 \sh^2\t +6)}
 \frac{-i}{2 \sqrt{2 \sh^2\t +6}} +\frac{-i}{12 \sqrt{6} })
 \times (\frac{1}{2} \b^2) \times 6.
\end{equation}

For type III diagram, there are two possible configurations.
For $b=\phi_1$, $c=\phi_2$, when $ k_1=0, k_2=\bar{k}$,
\begin{equation}
\bar{w}_1=\sqrt{2},~~\tilde{w}_1^+ = \sqrt{2} \ch\t +\sqrt{2 \sh^2\t+6},
 ~~ \tilde{w}_1^- =\sqrt{2} \ch\t -\sqrt{2 \sh^2\t+6},
\end{equation}
and when $k_1=\bar{k}, k_2=0$,
\begin{equation}
\bar{w}_1= \sqrt{2} \ch\t, ~~ \tilde{w}_1^+ = \sqrt{2} \ch\t+\sqrt{6},
  ~~ \tilde{w}_1^- =\sqrt{2} \ch\t-\sqrt{6}.
\end{equation}
For the symmetrized configuration of above, when $ k_1=0, k_2=\bar{k}$,
\begin{equation}
\bar{w}_1=\sqrt{6}, ~~ \tilde{w}_1^+ = 2 \sqrt{2} \ch\t,
 ~~ \tilde{w}_1^- =0,
\end{equation}
and when $k_1=\bar{k}, k_2=0$,
\begin{equation}
\bar{w}_1= \sqrt{2 \sh^2 \t +6},
  ~~ \tilde{w}_1^+ = \sqrt{2} \ch\t+\sqrt{2},
  ~~ \tilde{w}_1^- =\sqrt{2} \ch\t-\sqrt{2}.
\end{equation}
Inserting above data into Eq.(\ref{K-III}),
we obtain after some algebra,
\begin{eqnarray}
K_1^{(III-1)} &=& \frac{i \b^2}{4 \sqrt{2} \sh\t}
  (\frac{1}{4 \sqrt{2} (2 \ch\t +1)} + \frac{1}{12 \sqrt{2} \ch\t}
   - \frac{1}{12 \sqrt{2} (2\ch\t+\sqrt{3})}  \non \\
   &-& \frac{1}{4 \sqrt{2} (2 \ch\t-1)}
   +  \frac{1}{12 \sqrt{2} (2 \ch \t -\sqrt{3} )}
   - \frac{1}{ 4 \sqrt{2} ch\t (4 \ch^2 \t-3)}  \non \\
   &+& \frac{4 ch^2 \t+2}{ 2\sqrt{ 2 \sh^2 \t+6} (8 \ch^2 \t -2)} ).
\end{eqnarray}

For $b=\phi_3$, $c=\phi_4$, when $ k_1=0, k_2=\bar{k}$,
\begin{equation}
\bar{w}_1=\sqrt{2}, ~~ \tilde{w}_1^+ = 2 \sqrt{2} \ch\t,
  ~~ \tilde{w}_1^- =0,
\end{equation}
and when $k_1=\bar{k}, k_2=0$,
\begin{equation}
\bar{w}_1= \sqrt{2} \ch\t, ~~ \tilde{w}_1^+ = \sqrt{2} \ch\t+\sqrt{2},
  ~~ \tilde{w}_1^- =\sqrt{2} \ch\t-\sqrt{2}.
\end{equation}
Inserting above data into Eq.(\ref{K-III}),
we obtain after some algebra,
\begin{equation}
K_1^{(III-2)}= \frac{i \b^2}{4 \sqrt{2} \sh\t}
  (\frac{1}{ (2 \ch\t -1)}
   -\frac{1}{ ch\t (4 \ch^2 \t-1)}
   +\frac{1}{ \ch\t}
   - \frac{1}{ (2 \ch \t +1) } ) \frac{1}{2 \sqrt{2}}.
\end{equation}

Adding the above contributions as well as the tree result 1, we get
\begin{equation}
K_1(\t) = 1+ \frac{i \b^2}{24} (\frac{\sh\t}{ \ch\t}
           -\frac{\sh\t}{\ch\t-1}+\frac{2 \sh\t}{2 \ch\t +\sqrt{3}}
           -\frac{2 \sh\t}{2 \ch\t-1} ) +O(\b^4).
\label{K1}
\end{equation}
The non-meromorphic terms exactly cancel among themselves.

Now, we consider the heavy particle.
For type I diagram, there are four possible configurations
three of which yield identical contribution.
For $b= \phi_1, \phi_3$ and $\phi_4$,
\begin{equation}
K_2^{(I-1)}= \frac{1}{4 \sqrt{6} \sh\t} (\frac{1}{2 \sqrt{6 \sh^2\t+2} }
 +\frac{1}{2 \sqrt{2} })
    \times (\frac{-i \b^2}{4}) \times 2 \times 3.
\end{equation}
For $b=\phi_2$,
\begin{equation}
K_2^{(I-2)}= \frac{1}{4 \sqrt{6} \sh\t} ( \frac{1}{2 \sqrt{6} \ch\t }
 +\frac{1}{2 \sqrt{6} })
    \times ( \frac{-3 i}{8} \b^2) \times 12.
\end{equation}

For type II diagram, there are also four possible configurations
three of which yield identical contribution.
For $b=\phi_2$, $c=\phi_1, \phi_3$ and $\phi_4$,
\begin{equation}
K_2^{(II-1)}= \frac{1}{4 \sqrt{6} \sh\t}  (\frac{1}{(24 \sh^2\t +6)}
 \frac{-i}{2 \sqrt{6 \sh^2 \t+2}} +\frac{-i}{12 \sqrt{2} })
 \times (\frac{1}{2} \b^2) \times 6 \times 3.
\end{equation}
For $b=c=\phi_2$,
\begin{equation}
K_2^{(II-2)}= \frac{1}{4 \sqrt{6} \sh\t}  (\frac{1}{(24 \sh^2\t +6)}
 \frac{-i}{2 \sqrt{6} \ch \t} +\frac{-i}{12 \sqrt{6} })
 \times (\frac{-1}{2} \b^2) \times 18.
\end{equation}

For type III diagram, there are four possible configurations
three of which yield identical contributions.
For $b=c=\phi_1, \phi_3$ and $\phi_4$, when $ k_1=0, k_2=\bar{k}$,
\begin{equation}
\bar{w}_1=\sqrt{2},~~\tilde{w}_1^+ = \sqrt{6} \ch\t +\sqrt{6 \sh^2\t+2},
 ~~ \tilde{w}_1^- =\sqrt{6} \ch\t -\sqrt{6 \sh^2\t+2},
\end{equation}
and when $k_1=\bar{k}, k_2=0$,
\begin{equation}
\bar{w}_1= \sqrt{6 \sh^2 \t+2},
 ~~ \tilde{w}_1^+ = \sqrt{6} \ch\t+\sqrt{2},
  ~~ \tilde{w}_1^- =\sqrt{6} \ch\t-\sqrt{2}.
\end{equation}
Inserting above data into Eq.(\ref{K-III}),
we obtain after some algebra,
\begin{small}
\begin{equation}
K_2^{(III-1)}= \frac{3 i \b^2}{4 \sqrt{6} \sh\t}
  ( \frac{-1}{4 \sqrt{6} (2 \ch\t +\sqrt{3})}
 + \frac{1}{4 \sqrt{6} (2 \ch\t- \sqrt{3})}
 + \frac{ (2 ch^2 \t-1) }{ 2\sqrt{ 6 \sh^2 \t+2} (4 \ch^2 \t -3)} ).
\end{equation}
\end{small}

For $b=c=\phi_2$, when $ k_1=0, k_2=\bar{k}$,
\begin{equation}
\bar{w}_1=\sqrt{6}, ~~ \tilde{w}_1^+ = 2 \sqrt{6} \ch\t,
  ~~ \tilde{w}_1^- =0,
\end{equation}
and when $k_1=\bar{k}, k_2=0$,
\begin{equation}
\bar{w}_1= \sqrt{6} \ch\t, ~~ \tilde{w}_1^+ = \sqrt{6} \ch\t+\sqrt{6},
  ~~ \tilde{w}_1^- =\sqrt{6} \ch\t-\sqrt{6}.
\end{equation}
Inserting above data into Eq.(\ref{K-III}),
we obtain after some algebra,
\begin{equation}
K_2^{(III-2)}= \frac{9 i \b^2}{4 \sqrt{6} \sh\t}
  (\frac{1}{(2 \ch\t -1)}
   - \frac{1}{ ch\t (4 \ch^2 \t-1)}
   + \frac{1}{ \ch\t}
   - \frac{1}{ (2 \ch \t +1) } ) \frac{1}{12 \sqrt{6}}.
\end{equation}

Adding the above contributions as well as the tree result 1, we get
\begin{equation}
K_2(\t) = 1+ \frac{i \b^2}{24} (\frac{\sh\t}{ \ch\t}
           -\frac{\sh\t}{\ch\t-1}
           -\frac{4 \sh\t}{2 \ch\t -\sqrt{3}}
           +\frac{2 \sh\t}{2 \ch\t +\sqrt{3}}
           +\frac{2 \sh\t}{2 \ch\t+1} ) +O(\b^4).
\label{K2}
\end{equation}
Once again, the non-meromorphic terms exactly cancel among themselves.

This boundary reflection matrix up to one loop order satisfies
the boundary bootstrap equations up to $\b^2$ order.
\begin{eqnarray}
K_4(\t) &=& K_1(\t+i \pi 2/6) K_3(\t-i \pi 2/6)  S_{13}(2 \t),  \\
K_2(\t) &=& K_2(\t+i \pi 2/6) K_2(\t-i \pi 2/6)  S_{22}(2 \t), \non \\
K_2(\t) &=& K_1(\t+i \pi /6)  K_1(\t-i \pi /6)   S_{11}(2 \t),  \non
\end{eqnarray}
\begin{displaymath}
K_1(\t)=K_3(\t)=K_4(\t).
\end{displaymath}
If we consider all possible fusings as above,
the boundary crossing unitarity relations are automatically satisfied.

The exact boundary reflection matrix is determined uniquely
if we assume the strong-weak coupling \lq duality'.
\begin{eqnarray}
K_1(\t) &=& [1/2] [3/2] [5/2]^2 [7/2] [9/2], \\
K_2(\t) &=& [1/2] [3/2]^3 [5/2]^3 [7/2]^2 [9/2],  \non
\end{eqnarray}
where
\begin{equation}
 [ x ] = \frac{ (x-1/2) (x+1/2)} {(x-1/2+B/2) (x+1/2-B/2)}.
\end{equation}

On the other hand, the most general solution can be written in the
following form under the assumption of the strong-weak coupling
\lq duality'.
\begin{equation}
K_1(\t)= [1/2]^{a_1} [3/2]^{b_1} [5/2]^{c_1}
         [7/2]^{d_1} [9/2]^{e_1} [11/2]^{f_1},
\end{equation}
\begin{displaymath}
K_2(\t)= [1/2]^{a_2} [3/2]^{b_2} [5/2]^{c_2}
         [7/2]^{d_2} [9/2]^{e_2} [11/2]^{f_2}.
\end{displaymath}
Inserting the above into the boundary bootstrap equations, we can obtain
linear algebraic relations among the exponents. Solving this system of
equations yields the following result.
\begin{equation}
\begin{array}{lll}
 a_1 = free,        & b_1 = free,         & c_1 =a_1+b_1,      \\
 d_1 =a_1+b_1-1,    & e_1 =b_1,           & f_1 =a_1-1,     \non \\
 a_2 = -a_1+b_1+1,  & b_2 = 2 a_1 +b_1,   & c_2 =a_1+2 b_1, \non \\
 d_2 =a_1+2 b_1-1,  & e_2 =2 a_1 +b_1 -2, & f_2 =-a_1 +b_1. \non
\end{array}
\end{equation}

\section*{IV. Conclusions}
In this paper, we computed the boundary reflection
matrix for $d_4^{(1)}$ affine Toda field theory up to one loop order
in order to demonstrate a remarkable cancellation of non-meromorphic
terms which are always present for each diagram when the model
has a particle spectrum with more than one mass.

Using this result, we also determined the exact boundary reflection matrix
under the assumption of the strong-weak coupling \lq duality',
which turned out to be \lq non-minimal'. We also presented the complete
set of solutions of the boundary bootstrap equations.
Finally, we remark that Feynman diagrams which have no singularity
for the theory on a full line produce a new type of singularities.

\section*{Acknowledgement}
One(JDK) of the authors would like to thank Ed Corrigan and Ryu Sasaki.
He is also grateful to Choonkyu Lee, Soonkwon Nam and Changrim Ahn
for encouragement.
This work was supported by Korea Science and Engineering Foundation
and in part by the University of Durham.


\newpage
\begin{thebibliography}{99}
\bibitem{ZZ}
Al.B. Zamolodchikov and A.B. Zamolodchikov, Ann. Phys. 120 (1979) 253.
\bibitem{AFZ}
A.E. Arinshtein, V.A. Fateev and A.B. Zamolodchikov, Phys. Lett. B 87
(1979) 389.
\bibitem{BCDS}
H.W. Braden, E. Corrigan, P.E. Dorey and R. Sasaki, Nucl. Phys. B 338
(1990) 689.
\bibitem{CM}
P. Christe and G. Mussardo, Int. J. Mod. Phys. A 5 (1990) 4581.
\bibitem{DGZ}
G.W. Delius, M.T. Grisaru and D. Zanon, Nucl. Phys. B 382 (1992) 365.
\bibitem{BCDS2}
H.W. Braden, E. Corrigan, P.E. Dorey and R. Sasaki, Nucl. Phys. B 356
(1991) 469.
\bibitem{BS}
H.W. Braden and R. Sasaki, Phys. Lett. B 255 (1991) 343;
Nucl. Phys. B 379 (1992) 377.
\bibitem{CKK}
H.S. Cho, J.D. Kim and I.G. Koh, J. Math. Phys. 33 (1992) 2889.
\bibitem{BCKKS}
H.W. Braden, H.S. Cho, J.D. Kim, I.G. Koh and R. Sasaki,
Prog. Theor. Phys. 88 (1992) 1205.
\bibitem{SZ}
R. Sasaki and F.P. Zen, Int. J. Mod. Phys. A 8 (1992) 115.
\bibitem{ELOP}
R.J. Eden, P.V. Landshoff, D.I. Olive and J.C. Polkinghorne,
The analytic S matrix, (Cambridge University Press 1966).
\bibitem{Che}
I.V. Cherednik, Theor. Math. Phys. 61 (1984) 977.
\bibitem{GZ}
S. Ghoshal and A.B. Zamolodchikov, Int. J. Mod. Phys. A 9 (1994) 3841;
Int. J. Mod. Phys. A 9 (1994) 4353.
\bibitem{FK}
A. Fring and R. K\"oberle Nucl. Phys. B 421 (1994) 159;
Nucl. Phys. B 419 (1994) 647.
\bibitem{Gh}
S. Ghoshal, Int. J. Mod. Phys. A 9 (1994) 4801, hep-th/9310188.
\bibitem{Sasaki}
R. Sasaki, YITP/U-93-33, hep-th/9311027
in {\it Interface between Physics and Mathematics},
eds. W. Nahm and J-M. Shen, World Scientific (1994) 201.
\bibitem{CDRS}
E. Corrigan, P.E. Dorey, R.H. Rietdijk and R. Sasaki, Phys. Lett. B 333
(1994) 83.
\bibitem{Kim}
J.D. Kim, ``Boundary Reflection Matrix in Perturbative Quantum Field Theory",
DTP/95-11, hep-th/9504018, to appear in Phys. Lett. B.
\bibitem{Sym}
K. Symanzik, Nucl. Phys. B 190 (1981) 1.
\bibitem{DD}
H.W. Diehl and S. Dietrich, Z. Phys. B 50 (1983) 117.
\bibitem{BM}
M. Benhamou and G. Mahoux, Nucl. Phys. B 305 (1988) 1.
\end{thebibliography}
\end{document}